\documentclass{article}
\usepackage{bbm}
\usepackage{amsfonts}
\usepackage{palatino}
\usepackage{textcomp}
\usepackage{color}
\usepackage{amsmath}
\usepackage{amssymb}
\usepackage{mathbbold}
\usepackage{bbold}
\usepackage{graphicx}
\usepackage{dsfont}
\usepackage{float}
\usepackage{epstopdf}
\usepackage{hyperref}
\usepackage{authblk}

\usepackage{cite}

\def\no{\nonumber}

\newcommand{\bra}[1]{{\left< {#1} \right|}}
\newcommand{\ket}[1]{{\left| {#1} \right>}}
\renewcommand{\theequation}{\arabic{section}.\arabic{equation}}

\begin{document}

\bibliographystyle{unsrt}

\def\a{\alpha}
\def\b{\beta}
\def\d{\delta}
\def\e{\epsilon}
\def\g{\gamma}
\def\h{\mathfrak{h}}
\def\k{\kappa}
\def\l{\lambda}
\def\o{\omega}
\def\p{\wp}
\def\r{\rho}
\def\t{\tau}
\def\s{\sigma}
\def\z{\zeta}
\def\x{\xi}
\def\V={{{\bf\rm{V}}}}
 \def\A{{\cal{A}}}
 \def\B{{\cal{B}}}
 \def\C{{\cal{C}}}
 \def\D{{\cal{D}}}
\def\K{{\cal{K}}}
\def\O{\Omega}
\def\R{\bar{R}}
\def\T{{\cal{T}}}
\def\L{\Lambda}
\def\f{E_{\tau,\eta}(sl_2)}
\def\E{E_{\tau,\eta}(sl_n)}
\def\Zb{\mathbb{Z}}
\def\Cb{\mathbb{C}}

\def\R{\overline{R}}

\def\beq{\begin{equation}}
\def\eeq{\end{equation}}
\def\bea{\begin{eqnarray}}
\def\eea{\end{eqnarray}}
\def\ba{\begin{array}}
\def\ea{\end{array}}
\def\no{\nonumber}
\def\le{\langle}
\def\re{\rangle}
\def\lt{\left}
\def\rt{\right}

\newtheorem{Theorem}{Theorem}
\newtheorem{Definition}{Definition}
\newtheorem{Proposition}{Proposition}
\newtheorem{Lemma}{Lemma}
\newtheorem{Corollary}{Corollary}
\newcommand{\proof}[1]{{\bf Proof. }
        #1\begin{flushright}$\Box$\end{flushright}}

\title{Can a spin chain relate combinatorics to number theory?\\
\vspace{2mm}
\large Exact solution of the quantum integrable model \\
associated with the Motzkin spin chain
}

\author[1,2,3]{\large \bf Kun Hao\footnote{haoke72@163.com}}
\author[3]{\bf Olof Salberger}
\author[3,4]{\large \bf Vladimir Korepin\footnote{vladimir.korepin@stonybrook.edu}}

\affil[1]{\normalsize Institute of Modern Physics, Northwest University, Xi'an 710127, China}
\affil[2]{\normalsize Peng Huanwu Center for Fundamental Theory, Xi'an 710127, China}
\affil[3]{\normalsize C.N. Yang Institute for Theoretical Physics, Stony Brook University, NY 11794, USA}
\affil[4]{\normalsize Department of Physics and Astronomy, Stony Brook University, NY 11794, USA}

\maketitle
\begin{abstract}
The Motzkin spin chain is a spin-$1$ frustration-free model introduced by Shor \& Movassagh.
The ground state  is constructed by mapping random walks on the upper half of the square lattice  to spin configurations. It has unusually large entanglement entropy [quantum fluctuations].
The ground state of the Motzkin chain can be analytically described by the Motzkin paths. There is no analytical description of the excited states. The model is not solvable.
We simplify the model by removing one of the local equivalence moves of the Motzkin paths. The system becomes  integrable [similar to the XXX spin chain]. We call it free Motzkin chain.
From the point of view of quantum integrability, the model is special since its $R$-matrix does not have crossing unitarity.
We solve the periodic free Motzkin chain by generalizing the functional Bethe Ansatz method.
We construct a $T-Q$ relation with an additional parameter to formulate the energy spectrum.
This new parameter is related to the roots of unity and can be described by the M\"obius function in number theory. We observe further patterns of number theory.

\end{abstract}

\section{Introduction}
\setcounter{equation}{0}
Entanglement is a primary feature of quantum mechanics lacking in classical mechanics.
It can be used for building revolutionary new technologies (e.g., quantum computer) and information processing.
One way of measuring this quantity is by computing entanglement entropy (EE).
In this regard quantum many-body systems described by local Hamiltonians gain importance.
Restricting ourselves to $1D$ systems we know that gapped systems obey the area law \cite{Hastings2007,Alba20}.
The area law says that EE between two subsystems of a system is proportional to the area of the boundary between them. This is expected given the local nature of the interactions.
It is widely believed that for ``physically reasonable'' quantum systems, the area law can be violated by at most a logarithm $\log(N)$ factor\cite{Swingle13}, with $N$ the size of the quantum system.
This belief has been seriously challenged by both quantum information and condensed matter theorists in recent years.
It is of interest to find local Hamiltonians that exhibit more quantum fluctuations with their EE scaling as either the logarithm or the volume of the system size \cite{Irani10,Bravyi12}.

More recently, a spin-$1$ local, frustration-free Hamiltonian called the Motzkin spin chain \cite{Bravyi12,Movassagh16,DellAnna16} (Its half-integer analog is called the Fredkin spin chain \cite{Salberger18,Udagawa2017}) was shown to have unique ground states.
The model is constructed based on random walks, called the Motzkin walks.
Each Motzkin walk (path) can be mapped to a wave function of spins.
The Hamiltonian of the Motzkin chain is the nearest neighbors with projectors as interactions.
It has unique ground state that can be seen as a uniform superposition of Motzkin walks. As the volume of the system $N\rightarrow\infty$, the random Motzkin walks converge to Brownian excursions \cite{Durrett77}.
The EE of the system is proportional to $\sqrt{N}$.
It is exponentially more than the logarithm
\footnote{Actually, there is an infinite class of integer spin chain models that are physically reasonable and exact calculation of the EE shows that they violate the area law to the leading order by $\sqrt{N}$.}.

Such a spin chain can in principle be experimentally realized. Hydrodynamics of the model was studied in \cite{Richter2022}.
The large amount of entanglement may be used as a resource for quantum technologies and computation.
The model has since then been deformed to show phase transitions \cite{Salberger17,Zhang17}, generalized to possess new features using symmetric inverse semigroups \cite{Sugino18,Padmanabhan19} and simplified to preserve translational invariance in \cite{Caha18}.
The ground state of the Motzkin chain can be analytically expressed as a matrix product state of the Motzkin walks using techniques from enumerative combinatorics \cite{Stanley1999}.
While there is no analytical description of the excited states, the model is not exactly solvable.

The full Hamiltonian of the uncolored\footnote{The color was introduced  by Shor \& Movassagh \cite{Movassagh16}.} Motzkin chain was described by Bravyi et al. \cite{Bravyi12}.
It is a spin chain with a 3-dimensional local Hilbert space given by a basis $|{u}\rangle$, $|{f}\rangle$, $|{d}\rangle$. Here $u$, $f$, and $d$ are used to abbreviate ``up'', ``flat'', and ``down'' of the Motzkin walks respectively.
This identification comes from that the states in this system can be mapped to paths (Motzkin walks) in the ``x-y'' plane as done in \cite{Movassagh16}.
The Hamiltonian of the Motzkin spin chain with periodic boundary conditions is given by (the coupling constant $g$ was equal to $1$ in \cite{Movassagh16}):
\begin{equation}
 H_{M} = \sum_{j} H_{free,j} + g H_{int,j}
\end{equation}

The local Hamiltonians $H_{free}$ and $H_{int}$ are given in terms of two-site projectors,
\begin{eqnarray}
H_{free,j}=(|{u_jf_{j+1}}\rangle-|{f_ju_{j+1}}\rangle) (\langle{u_jf_{j+1}}|-\langle{f_ju_{j+1}}|)
+(|{d_jf_{j+1}}\rangle -|{f_jd_{j+1}}\rangle)(\langle{d_jf_{j+1}}|- \langle{f_jd_{j+1}}|)
\end{eqnarray}
and
\begin{equation}
H_{int,j} =(|{u_jd_{j+1}}\rangle - |{f_jf_{j+1}}\rangle) (\langle{u_jd_{j+1}}| - \langle{f_jf_{j+1}}|).
\end{equation}
The term $H_{int}$ can be viewed as a ``kinetic part'' of the Motzkin model.
By turning on $g$, the system loses its symmetries and starts to ``interact''.
Only the ground state of the chain can be analytically described by the Motzkin paths. The model is not exactly solvable.
Thus we will call the case with $g=0$ the non-interacting Motzkin chain or the free Motzkin spin chain.
While it is still strongly interacting, it is indeed integrable in the sense of quantum inverse scattering method \cite{Faddeev1995,Korepin93} (or the algebraic Bethe Ansatz method (ABA)), as expanded on below. This fact allows us to solve the spectrum of the excited states completely.

In this paper we will analyze the case of the free Motzkin spin chain
\begin{equation}
H_{FM}=H_{M}|_{g=0}= \sum_{j} H_{free,j}
\end{equation}
We observe that the resulting system is integrable. Its structure can be described (generated) by Yang-Baxter equation (Eq.\ref{Yang-Baxter-equation}). Then we will calculate its spectrum for periodic boundary conditions\footnote{The integrability of the model with open boundaries was discussed in \cite{Tong2021}.}.
The model can be seen as a generalized Temperley-Lieb system.
This is understood from the local Hamiltonian, which is a representation of the generator of Temperley-Lieb algebra.
Due to the lack of commutation relations among the monodromy-matrix operators(Eq.\ref{T-matrix}), it is hard to construct the Bethe state and further steps of the standard ABA.
We generalize Sklyanin's functional Bethe Ansatz \cite{Sklyanin92,Sklyanin95} and some ideas of Off-diagonal Bethe Ansatz \cite{Wang2015,Cao2013} to approach the spectrum problem of the model\footnote{The method is especially capable of solving integrable models without (reference state and Bethe state) $U(1)$ symmetry by introducing a inhomogeneous term in $T-Q$ relation. In the current work, we also skip the step of constructing the Bethe state.}.

This paper is organized as follows. In section 2, we introduce the integrable Hamiltonian and its associated integrable structure.
In section 3, we briefly review the results of XXX spin-$1\over2$ chain case. In section 4, in contrast with the XXX spin-$1\over2$ chain case, we derive certain operator product identities for the transfer matrix of the model by using some intrinsic properties of the $R$-matrix. The asymptotic behavior of the transfer matrix are also obtained. Section 5 is devoted to the construction of the $T-Q$ relation and the corresponding Bethe Ansatz equations.
In section 6, we summarize our results and give some discussions.
Some detailed technical derivation is given in Appendixes.

\section{The Free Motzkin spin chain as an integrable model}
\setcounter{equation}{0}
To readers familiar with other similar models but not with the Bethe Ansatz method, the central building block is the $R$-matrix. It essentially describes the scattering between two quasiparticles in the spin chain. The $R$-matrix satisfies the Yang-Baxter equation, which states that many-body scattering can be reduced to two-body scattering. This fact ensures the integrability of the periodic free Motzkin chain.

The $R$-matrix $R_{12}(\lambda)=P_{12}[(\lambda+\eta)\mathbbold{1}-\lambda\hat{e}_{12}]$ for the free Motzkin spin chain can be written as
\begin{equation}
\label{R-matrix-Motzkin}
\left(
\begin{array}{ccc|ccc|ccc}
\lambda+\eta &    &    &    &    &    &    &    &   \\
      &\lambda&    &\eta&    &    &    &    &   \\
      &    &    &    &    &    &\lambda+\eta&    &   \\
\hline
      &\eta&    &\lambda&    &    &    &    &   \\
      &    &    &    &\lambda+\eta&    &    &    &   \\
      &    &    &    &    &\lambda&    &\eta&   \\
\hline
      &    &\lambda+\eta&    &    &    &    &    &    \\
      &    &    &    &    &\eta&    &\lambda&    \\
      &    &    &    &    &    &    &    &\lambda+\eta\\
\end{array}
\right).
\end{equation}
Here $\lambda$ is the spectral parameter. $\eta$ is the crossing parameter. $P_{12}$ is the permutation operator, $\hat{e}_{12}$ is the corresponding generator \cite{Temperley-Lieb} of the Temperley-Lieb algebra, and ${1\over2}\hat{e}_{12}$ is a projection operator.
\begin{equation}
\hat{e}_{12}=\hat{U}_{12}+\hat{D}_{12}.
\end{equation}
Unless otherwise specified, we use the standard notation: for any matrix $A\in {\rm End}({\rm\bf V})$, $A_j$ is an embedding operator in the tensor space ${\rm\bf V}\otimes
{\rm\bf V}\otimes\cdots$. It acts as $A$ on the $j$-th space and as an identity on the other factor spaces. So $R_{ij}(\lambda)$ acts as an identity on the factor spaces except for the $i$-th and $j$-th ones.
We denote the basis states by $\{|{u}\rangle, |{f}\rangle, |{d}\rangle\}$ where $u$, $f$ and $d$ are used to abbreviate ``up'', ``flat'' and ``down'' respectively,
\begin{equation}
|{u}\rangle=\left(\begin{array}{c}
    1 \\
    0 \\
    0 \\
  \end{array}\right),\quad
|{f}\rangle=\left(\begin{array}{c}
    0 \\
    1 \\
    0 \\
  \end{array}\right),\quad
|{d}\rangle=\left(\begin{array}{c}
    0 \\
    0 \\
    1 \\
  \end{array}\right).
\end{equation}
The operators $\hat{U}_{j,j+1}$ and $\hat{D}_{j,j+1}$ are projectors to the states
$|{u_j,f_{j+1}}\rangle\hspace{-0.5mm}-\hspace{-0.5mm}|{f_j,u_{j+1}}\rangle,\;\text{and}\;\,|{d_j,f_{j+1}}\rangle\hspace{-0.5mm}- \hspace{-0.5mm}|{f_j,d_{j+1}}\rangle$
respectively,
\begin{eqnarray}
\hat{U}_{j,j+1} = \ket{u_j,~f_{j+1}}\bra{u_j,~f_{j+1}} - \ket{u_j,~f_{j+1}}\bra{f_j,~u_{j+1}} - \ket{f_j,~u_{j+1}}\bra{u_j,~f_{j+1}} + \ket{f_j,~u_{j+1}}\bra{f_j,~u_{j+1}},
\end{eqnarray}and
\begin{eqnarray}
\hat{D}_{j,j+1} =  \ket{d_j,~f_{j+1}}\bra{d_j,~f_{j+1}} - \ket{d_j,~f_{j+1}}\bra{f_j,~d_{j+1}} - \ket{f_j,~d_{j+1}}\bra{d_j,~f_{j+1}} + \ket{f_j,~d_{j+1}}\bra{f_j,~d_{j+1}}.
\end{eqnarray}
Alternatively, they also can be expressed in terms of standard basis and Pauli matrices like representation,
\begin{equation}
\hat{U}_{j,j+1} = E_j^{11}E_{j+1}^{22}-E_j^{12}E_{j+1}^{21}-E_j^{21}E_{j+1}^{12}+E_j^{22}E_{j+1}^{11} =
\frac{1}{2}\left[ I^u_jI^u_{j+1} - u^x_ju^x_{j+1}- u^y_ju^y_{j+1}- u^z_ju^z_{j+1}\right],
\end{equation}
with
\begin{equation}
I^u = \left(\begin{array}{ccc} 1 & 0 & 0 \\ 0 & 1 & 0 \\ 0 & 0 & 0\end{array}\right),~ u^x = \left(\begin{array}{ccc} 0 & 1 & 0 \\ 1 & 0 & 0 \\ 0 & 0 & 0\end{array}\right), ~ u^y = \left(\begin{array}{ccc} 0 & -i & 0 \\ i & 0 & 0 \\ 0 & 0 & 0\end{array}\right), ~ u^z = \left(\begin{array}{ccc} 1 & 0 & 0 \\ 0 & -1 & 0 \\ 0 & 0 & 0\end{array}\right).
\label{def-u}\end{equation}
In a similar manner, we have the expression for operator $\hat{D}_{j,j+1}$,
\begin{equation}
\hat{D}_{j,j+1} = E_j^{33}E_{j+1}^{22}-E_j^{32}E_{j+1}^{23}-E_j^{23}E_{j+1}^{32}+E_j^{22}E_{j+1}^{33}=
\frac{1}{2}\left[ I^d_jI^d_{j+1} - d^x_jd^x_{j+1}- d^y_jd^y_{j+1}- d^z_jd^z_{j+1}\right],
\end{equation}
with
\begin{equation}
I^d = \left(\begin{array}{ccc} 0 & 0 & 0 \\ 0 & 1 & 0 \\ 0 & 0 & 1\end{array}\right),~ d^x = \left(\begin{array}{ccc} 0 & 0 & 0 \\ 0 & 0 & 1 \\ 0 & 1 & 0\end{array}\right), ~ d^y = \left(\begin{array}{ccc} 0 & 0 & 0 \\ 0 & 0 & -i \\ 0 & i & 0\end{array}\right), ~ d^z = \left(\begin{array}{ccc} 0 & 0 & 0 \\ 0 & 1 & 0 \\ 0 & 0 & -1\end{array}\right).
\label{def-d}\end{equation}

The $R$-matrix satisfies the quantum Yang-Baxter equation (QYBE)
\begin{equation}
R_{12}(\lambda)R_{13}(\lambda+\nu)R_{23}(\nu)=R_{23}(\nu)R_{13}(\lambda+\nu)R_{12}(\lambda),
\label{Yang-Baxter-equation}\end{equation}
and possesses the following properties:
\begin{eqnarray}
&\mbox{Initial condition:}~~~~~~ R_{12}(0)= \eta P_{12},~~~~~\label{Initial-R}\\
&\mbox{Unitarity:}~  R_{12}(\lambda)R_{21}(-\lambda)= (\eta+\lambda)(\eta-\lambda)\,\mathbbold{1},\label{Unitarity}\\
&\mbox{Projection:}~~~  R_{12}(-\eta)=\eta P_{12}\hat{e}_{12}=- \eta\hat{e}_{12}.\label{Projection}
\end{eqnarray}
The partial transpose of this $R$-matrix is degenerate
\begin{equation}
\det(R^{t_1}_{12}(\lambda))=0.
\label{partial-trans}
\end{equation}
Therefore it does not satisfy crossing unitarity\cite{Perk06}. It can be seen as a special solution of Yang-Baxter equation\cite{Perk1990}.
This fact also makes the standard ABA procedure for Temperley-Lieb algebra \cite{Nepomechie2016} does not work on this model.
The monodromy-matrix $T(\lambda)$ can be constructed as
\begin{equation}
T(\lambda)=R_{0N}(\lambda-\theta_N)R_{0\,N-1}(\lambda-\theta_{N-1})\cdots R_{01}(\lambda-\theta_1),
\label{Monodromy-T}\end{equation}
where the subscript $0$ indicates the auxiliary space and the indices $\{1, \cdots, N\}$ denote
the physical or quantum spaces. $N$ is the number of sites and $\{\theta_j|j=1, \cdots, N\}$ are the inhomogeneous parameters.

Then the monodromy-matrix can be written as
\begin{equation}
\label{T-matrix}
T_0(\lambda)=
\left(
\begin{array}{ccc}
A(\lambda) & B_1(\lambda) & B_2(\lambda)\\
C_1(\lambda)&D_{11}(\lambda)&D_{12}(\lambda)\\
C_2(\lambda)&D_{21}(\lambda)&D_{22}(\lambda)\\
\end{array}
\right).
\end{equation}
Here the subscripts of $B$, $C$, $D$ operators indicate their position in auxiliary space $0$.
The transfer matrix $t(\lambda)$ is defined as the partial trace of monodromy matrix in the auxiliary
space
\begin{equation}
t(\lambda) = tr_0\{T_0(\lambda)\}=A(\lambda)+D_{11}(\lambda)+D_{22}(\lambda).
\label{trans}\end{equation}
Let us further introduce the vacuum state $|0\rangle$,
\begin{equation}
|0\rangle=|u_1\rangle,\cdots,|u_N\rangle=\left(\begin{array}{c}
    1 \\
    0 \\
    0 \\
  \end{array}\right)_{1} \otimes\cdots\otimes\left(\begin{array}{c}
    1 \\
    0 \\
    0 \\
  \end{array}\right)_{N} .
\end{equation}
The operators acting on the vacuum state give rise to
\begin{eqnarray}
&&A(\lambda)|0\rangle=a(\lambda)|0\rangle=\prod^N_{j=1}(\lambda-\theta_j+\eta)|0\rangle,\nonumber\\ &&D_{11}(\lambda)|0\rangle=d(\lambda)|0\rangle=\prod^N_{j=1}(\lambda-\theta_j)|0\rangle,\nonumber\\ &&D_{22}(\lambda)|0\rangle=0\quad,C_k(\lambda)\,|0\rangle=0,\quad B_k(\lambda)\,|0\rangle\neq 0.
\label{A-D-eigen}\end{eqnarray}
The QYBE (Eq.\ref{Yang-Baxter-equation}) implies that the monodromy-matrix $T(\lambda)$ satisfies the Yang-Baxter relation
\begin{equation}
R_{12}(\lambda-\nu)T_{1}(\lambda)T_{2}(\nu)=T_{2}(\nu)T_{1}(\lambda)R_{12}(\lambda-\nu).
\end{equation}
Therefore, one can prove that the transfer matrices with different
spectral parameters commute with each other
\begin{eqnarray}
[t(\lambda),t(\nu)]=0.
\label{trans-com}
\end{eqnarray}
So $t(\lambda)$ serves as the generating functional of the conserved quantities of the system.
This ensures the integrability of the periodic free Motzkin spin chain.

The Hamiltonian of periodic free Motzkin spin chain can be recovered from the transfer matrix through a logarithmic trace identity, such that
\begin{eqnarray}
H_{FM}\hspace{-0.5mm}=\hspace{-0.5mm}-\eta\;\frac{\partial\ln t(\lambda)}{\partial \lambda}\bigg{|}_{\lambda=0,\{\theta_j=0\}}\hspace{-0.5mm}+N=\sum^{N}_{j=1}H_{j,j+1}.~
\label{Hamiltonian}\end{eqnarray}
Here the local Hamiltonian is
\begin{eqnarray}
H_{j,j+1}=H_{free,j}=\hat{e}_{j,j+1}=\hat{U}_{j,j+1}+\hat{D}_{j,j+1}\;,
\end{eqnarray}
with the periodic boundary conditions
$H_{N,N+1}=H_{N,1}$.

\section{A brief review of XXX spin-$1\over2$ chain}\label{sec:review XXX spin-1/2}
\setcounter{equation}{0}
The final results show that the spectrum of the free Motzkin chain can be constructed based on the XXX spin-$1\over2$ chain. Here we take the XXX spin-$1\over2$ chain as an example to further explain the situation in the free Motzkin chain. Following the concept of functional Bethe Ansatz or Separation of Variables (SoV) method, we have the operator identities for transfer matrix of XXX spin-$1\over2$ chain\footnote{The superscript $({1\over2})$ refers to spin-$1\over2$, and the functions $a(\lambda)$ and $d(\lambda)$ are the same as that of free Motzkin case.},
\begin{eqnarray}
t^{({1\over2})}\hspace{-0.5mm}(\theta_j)t^{({1\over2})}\hspace{-0.5mm}(\theta_j-\eta)\hspace{-1mm}=\hspace{-1mm}\mathbbold{1}\cdot a(\theta_j)d(\theta_j-\eta),\; j\hspace{-1mm}=\hspace{-1mm}1,\hspace{-0.5mm}\cdots\hspace{-0.5mm},\hspace{-0.5mm}N\hspace{-0.5mm}.~~
\label{trans-id-XXX}\end{eqnarray}
Next, we consider the asymptotic behaviors of fused transfer matrix
\begin{eqnarray}
\lim_{\lambda\to \infty} t^{({1\over2})}(\lambda)=\mathbbold{1}\cdot(2\lambda^N+\cdots).
\label{Asym-XXX}\end{eqnarray}
Then let us assume the eigenstate of $ t^{({1\over2})}(\lambda)$ is $|\psi\rangle$, with its eigenvalue $\Lambda^{({1\over2})}(\lambda)$. Acting the above operator identities (Eq.\ref{trans-id-XXX}) and the asymptotic relation (Eq.\ref{Asym-XXX}) on $|\psi\rangle$, they should satisfy the corresponding functional relations\footnote{People mainly focused on functional relations in previous works, since the transition between the operator identities and functional relations is straightforward. Also, the calculation of the functional relations is relatively easy. However, in our current model, we should emphasize the operator identities. See the next section for details.}
\begin{eqnarray}
\Lambda\hspace{-0.5mm}^{({1\over2})}\hspace{-0.5mm}(\theta_j)\Lambda\hspace{-0.5mm}^{({1\over2})}\hspace{-0.5mm}(\theta_j-\eta)\hspace{-1mm}=\hspace{-1mm}a(\theta_j)d(\theta_j-\eta),\; j\hspace{-0.5mm}=\hspace{-0.5mm}1,\hspace{-0.5mm}\cdots\hspace{-0.5mm},\hspace{-0.5mm}N.~~
\end{eqnarray}\begin{eqnarray}
\lim_{\lambda\to \infty} \Lambda^{({1\over2})}(\lambda)=2\lambda^N+\cdots
\end{eqnarray}
As we know, the $T-Q$ relation of the XXX spin-$1\over2$ chain is a polynomial function of degree $N$. This can be seen from the construction of the monodromy matrix.
The above $N+1$ functional relations are sufficient to determine such a $T-Q$ relation with Bethe equations
\begin{eqnarray}
\Lambda^{({1\over2})}(\lambda)=a(\lambda)\frac{Q(\lambda-\eta)}{Q(\lambda)}+d(\lambda)\frac{Q(\lambda+\eta)}{Q(\lambda)}.
\label{T-Q-XXX}
\end{eqnarray}
The function $Q(\lambda)$ is parameterized by $N$ Bethe roots $\{\lambda_j|j=1,\cdots,N\}$ as follows\footnote{We remark that the form of function $Q(\lambda)$ here is the same as that of the free Motzkin case in a later section.}:
\begin{eqnarray}
Q(\lambda)=\prod^{N}_{j=1}(\lambda-\lambda_j).
\end{eqnarray}
To ensure $\Lambda^{({1\over2})}(\lambda)$ to be a polynomial, the residues at the
poles $\lambda_j$ must vanish. This requires the $N$ Bethe roots to satisfy the Bethe equations
\begin{eqnarray}
\prod^N_{k=1}\frac{\lambda_j-\theta_k+\eta}{\lambda_j-\theta_k}
\hspace{-1mm}=\hspace{-1mm}\prod^N_{l\neq j\atop{l=1}}\frac{\lambda_j-\lambda_l+\eta}{\lambda_j-\lambda_l-\eta},\;\, j=1,\hspace{-0.5mm}\cdots\hspace{-0.5mm},\hspace{-0.5mm}N.~~
\end{eqnarray}

\section{Operator identity of the transfer matrix}
\setcounter{equation}{0}
For the current free Motzkin chain model, it is difficult to implement the standard ABA method. Let us go along with the concept of functional BA \cite{Sklyanin92,Sklyanin95} and consider the products of the transfer matrices at special points.
To do this, we follow the method in \cite{Wang2015} and take advantage of the symmetry and properties (Eq.\ref{Initial-R}), (Eq.\ref{Unitarity}) and (Eq.\ref{Projection}) of the $R$-matrix. After a long calculation, we
find that the transfer matrix (Eq.\ref{trans}) satisfies the following operator identities
\begin{eqnarray}
t(\theta_j)t(\theta_j-\eta)=\hat{O}a(\theta_j)d(\theta_j-\eta),\;\, j=1,\cdots,N.~
\label{Operator-identity}\end{eqnarray}
Other than the identity operator (matrix) on the r.h.s of (Eq.\ref{trans-id-XXX}) 
for the spin-$1\over2$ chain case, here the operator $\hat{O}$ is a special operator.
Thus we can not simply convert the operator identities to the functional relations like in the XXX spin-$1\over2$ chain and all other cases. The final results show that $\hat{O}$ is a global operator acting highly non-trivial on all quantum spaces. One can not derive the functional relation by projecting the transfer matrix to its subspace. The fusion techniques \cite{Klumper1992} do not work here.

So for the current model, the key point is to understand the structure of the $\hat{O}$ operator.
The form of operator $\hat{O}$ consists of the leading terms of operators $A(\lambda)$, $D_{11}(\lambda)$ and $D_{22}(\lambda)$ as polynomials of variable $\lambda$. It can be expressed as follows
\begin{equation}
\hat{O}={\displaystyle \lim_{\lambda \to \infty}{1\over\lambda^{2N}}(A(\lambda)+D_{22}(\lambda))D_{11}(\lambda)}
={\displaystyle \lim_{\lambda \to \infty}{1\over \lambda^N}(A(\lambda)+D_{22}(\lambda))}.
\label{O-operator}\end{equation}
The second step of the formula above results from the leading terms of operator $D_{11}(\lambda)$ is proportional to the identity operator (matrix).

Next, let us briefly analyze the asymptotic behaviors of operators $A(\lambda)$ and $D_{22}(\lambda)$ as $\lambda$ tends to infinity.
The operators $A(\lambda)$ and $D_{22}(\lambda)$ are centrally symmetric to each other with respect to the element $E^{{(3^N+1)/2},{(3^N+1)/2}}$ (the center element of the matrix acting on the full quantum spaces). This means each element $E^{i,j}$ (we adopt the standard basis) in operator $A(\lambda)$ can be mapped to the element $E^{3^N+1-i,3^N+1-j}$ in operator $D_{22}(\lambda)$ after a $180$-degree rotation around the central element. Here $3^N$ is the dimension of $N$ quantum spaces.

The operators $A(\lambda)$ and $D_{22}(\lambda)$ have the following asymptotic behaviors as $\lambda$ goes to infinity.
The leading terms of $A(\lambda)$ operator is a degenerate operator acting on column vectors in the up half quantum space $\begin{bmatrix}x_{1},x_{2},\dots,x_{M},0,\dots,0\end{bmatrix}^{\rm {T}}$, and $M={3^N+1\over2}$.
This degenerate operator can be seen as a series of row transformations acting on this diagonal matrix
\begin{eqnarray}
\operatorname{diag}(1,\vspace{-0.5mm}\cdots\vspace{-0.5mm},1,0,\vspace{-0.5mm}\cdots\vspace{-0.5mm},0)\vspace{-0.5mm}=\vspace{-0.5mm}\sum^M_{j=1}1\vspace{-0.5mm}\times\vspace{-0.5mm}E^{j,j},\;\, M\vspace{-0.5mm}=\vspace{-0.5mm}{3^N\vspace{-0.5mm}+\vspace{-0.5mm}1\over2}.~~~~
\end{eqnarray}

Here, from the first diagonal element $E^{1,1}$ to the $M$-th diagonal element $E^{M,M}$, $M$ elements each have the value $1$, and the other elements are all $0$.

Considering the symmetry between $A(\lambda)$ and $D_{22}(\lambda)$, the leading terms of the composite operator $A(\lambda)+D_{22}(\lambda)$ is a non degenerate operator. It can be expressed as
a series of row transformations acting on the diagonal matrix
\begin{eqnarray}
\operatorname {diag} (1,\vspace{-0.5mm}\cdots\vspace{-0.5mm},1,2,1,\vspace{-0.5mm}\cdots\vspace{-0.5mm},1)\vspace{-0.5mm}=\vspace{-0.5mm}\sum^{3^N}_{j=1\atop j\neq M}1\vspace{-0.5mm}\times\vspace{-0.5mm}E^{j,j}\vspace{-0.5mm}+\vspace{-0.5mm}2\vspace{-0.5mm}\times\vspace{-0.5mm}E^{M,M}~~~~~
\end{eqnarray}
This is just an identity matrix with its center ($M$-th) diagonal element plus $1$.

Moreover, the asymptotic behavior of the transfer matrix is
\begin{equation}
\lim_{\lambda\to\infty}t(\lambda)=\lim_{\lambda\to\infty}A(\lambda)+D_{11}(\lambda)+D_{22}(\lambda)
=\lambda^N(\hat{O}+\mathbbold{1})+\cdots
\label{trans-asymptotic}\end{equation}
The operator identities (Eq.\ref{Operator-identity}) of the transfer matrix together with the asymptotic behavior (Eq.\ref{trans-asymptotic}) completely determine the spectrum of the transfer matrix.

The detailed analysis of the structure of operator $\hat{O}$ (also operators $A(\lambda)$ and $D_{22}(\lambda)$) and its eigenvalues are given in Appendix A.

\section{Functional relations and the $T-Q$ relation}
\setcounter{equation}{0}
The commutativity of the transfer matrices (Eq.\ref{trans-com}) means that they have common eigenstates. Let us assume that $|\Psi\rangle$ is an eigenstate of $t(\lambda)$. $|\Psi\rangle$ does not depend upon $\lambda$, and its corresponding eigenvalue is $\Lambda(\lambda)$,
\begin{eqnarray}
t(\lambda)|\Psi\rangle=\Lambda(\lambda)|\Psi\rangle.
\label{Lambda}\end{eqnarray}
The operator identities (Eq.\ref{Operator-identity}) imply that the corresponding eigenvalue $\Lambda(\lambda)$ satisfies the relation
\begin{eqnarray}
\Lambda(\theta_j)\Lambda(\theta_j-\eta)|\Psi\rangle=\hat{O}a(\theta_j)d(\theta_j-\eta)|\Psi\rangle
=\{\omega\} a(\theta_j)d(\theta_j-\eta)|\Psi\rangle,\quad j=1,\cdots,N.
\end{eqnarray}
Thus we have
\begin{eqnarray}
\Lambda(\theta_j)\Lambda(\theta_j-\eta)
\hspace{-0.5mm}=\hspace{-0.5mm}\{\omega\} a(\theta_j)d(\theta_j-\eta),\;\, j=1,\hspace{-0.5mm}\cdots\hspace{-0.5mm},N.~~
\label{Lambda-identity}\end{eqnarray}
Here $\{\omega\}$ represents the eigenvalues of operator $\hat{O}$, which have $3^N$ numbers, such as
\begin{eqnarray}
\{2,1,\cdots,-1,\cdots,i,-i,\exp({2\pi i\over3}),\exp({4\pi i\over3}),\cdots,
\exp({2\pi i\over N}),\exp({4\pi i\over N}),\cdots\}.
\label{eigen-O}\end{eqnarray}
The eigenvalues $\{\omega\}$ of operator $\hat{O}$ include two parts: The diagonal elements result in one single $2$ (the center element) and several $1$ (with degeneracy of $2^{N+1}-2$); The non-diagonal elements give rise to roots of unity, ranging from square roots to $N$-th roots.

In general, the distribution of $k$-th roots of unity as eigenvalues of operator $\hat{O}$ for a lattice of length $N$ is given by:
\begin{eqnarray}
f(N,k) = \left( \sum_{d|k} 2^d \mu(\frac{k}{d}) \right) \left[ \sum_{m = 1}^{\lfloor{\frac{N}{k}}\rfloor} C(N,km)
\right]
\label{Formula_Mobius}\end{eqnarray}
where $\mu(n)$ refers to the M\"obius function in number theory. $C(n,k)$ is the inline notation of the binomial coefficient $\binom {n}{k}$. For $k=1$ this formula gives the correct diagonal result if one excludes the eigenvalue $2$. The $1$ with degeneracy of $2^{N+1}-2$ from the diagonal elements can be regarded as the $1$-st roots of unity.
In every specific $k$-th roots case, each kind of roots $\exp({2\pi i\over k})$, $\exp({4\pi i\over k})$,$\cdots$, $1$ has the same degeneracy. See Appendix A and B for detailed derivation.

According to the asymptotic behavior of the transfer matrix (Eq.\ref{trans-asymptotic}), we have that the leading terms of $\Lambda(\lambda)$ have the following values,
\begin{eqnarray}
&&\lambda^N\hspace{-0.5mm}\times\hspace{-0.5mm}\{3,2,\cdots,0,\cdots,1+i,1-i,1+\exp({2\pi i\over3}),\nonumber\\
&&1+\exp({4\pi i\over3}),\hspace{-0.5mm}\cdots\hspace{-0.5mm},1+\exp({2\pi i\over N}),1+\exp({4\pi i\over N}),\hspace{-0.5mm}\cdots\hspace{-0.5mm}\}.~~~
\label{Lambda-asymptotic}\end{eqnarray}
They are $\lambda^N$ multiplied by each eigenvalue of operator $\hat{O}$ plus $1$, respectively. They have a one-to-one correspondence with the eigenvalues of operator $\hat{O}$ (Eq.\ref{eigen-O}).

$\Lambda(\lambda)$, as a function of $\lambda$, is a polynomial of degree $N$.
The functional relations (Eq.\ref{Lambda-identity}) together with this asymptotic behavior (Eq.\ref{Lambda-asymptotic}), these $N+1$ conditions are sufficient to construct the $T-Q$ relation \cite{Baxter82} for $\Lambda(\lambda)$.

Due to $\{\omega\}$, other than XXX spin-${1\over2}$ chain case, there must exist coefficients before the two terms of the $T-Q$ relation(Eq.\ref{T-Q}). Assuming $\omega$ is one eigenvalue of operator $\hat{O}$, then the corresponding coefficient of the leading term of $\Lambda(\lambda)$ as $\lambda\rightarrow\infty$ should be $\omega+1$, as shown in (Eq.\ref{Lambda-asymptotic}). Let $\tilde{a}$ and $\tilde{d}$ represent the coefficients before $a(\lambda)$ and $d(\lambda)$ terms, respectively.
Based on the identities (Eq.\ref{Lambda-identity}) and the asymptotic behavior (Eq.\ref{trans-asymptotic}) and (Eq.\ref{Lambda-asymptotic}), these coefficients should satisfy the following equations
\begin{eqnarray}
\left\{
\begin{aligned}
&\tilde{a}\cdot \tilde{d}=\omega,\\
&\tilde{a}+\tilde{d}=\omega+1.
\end{aligned}
\right.
\end{eqnarray}
The equations have two sets of solutions:
\begin{eqnarray}
\left\{
\begin{aligned}
&\tilde{a}=\omega,\\
&\tilde{d}=1;
\end{aligned}
\right.
\qquad\text{and}\qquad
\left\{
\begin{aligned}
&\tilde{a}=1,\\
&\tilde{d}=\omega.
\end{aligned}
\right.
\label{coefficient-solution}\end{eqnarray}
After carefully calculations, we find that only the first solution set is appropriate\footnote{The discussion is based on the eigenvalues (Eq.\ref{A-D-eigen}) of the operators $A(\lambda)$, $D_{11}(\lambda)$ and the structure of the operator $\hat{O}$ mentioned above.}.
Basically speaking, this is because $\{\omega\}$ are eigenvalues of the operator $\hat{O}$, and the operator $\hat{O}$ can be related to $A(\lambda)+D_{22}(\lambda)$, which correspond to $a(\lambda)$ (Eq.\ref{A-D-eigen}). In contrast, the operator $D_{11}(\lambda)$ corresponds to $d(\lambda)$, and $D_{11}(\lambda)$ is proportional to identity operator (matrix) under the limit $\lambda$ tends to $\infty$.

\subsection{$T-Q$ Ansatz and the spectrum}
To explain the situation in the free Motzkin chain, we have presented a quick review of how to apply the functional BA approach to the XXX spin-$1\over2$ chain in Section 3.
We propose a generalized $T-Q$ relation to describe the eigenvalues of the transfer matrix (Eq.\ref{trans}). This $T-Q$ relation, together with the newly appeared $\hat{O}$ operator (Eq.\ref{O-operator}) in the current model, allows us to calculate all spectrum of the Hamiltonian.

Based on the properties (Eq.\ref{Lambda-identity}), (Eq.\ref{Lambda-asymptotic}), and also the configuration of the coefficient $\omega$ (Eq.\ref{coefficient-solution}), let us propose the following conjecture for $\Lambda(\lambda)$,
\begin{eqnarray}
\Lambda(\lambda)=\omega\;a(\lambda)\frac{Q(\lambda-\eta)}{Q(\lambda)}+d(\lambda)\frac{Q(\lambda+\eta)}{Q(\lambda)}.
\label{T-Q}\end{eqnarray}
Here $\omega$ will go through each value in the set $\{\omega\}$. The function $Q(\lambda)$ is parameterized by $N$ Bethe roots $\{\lambda_j|j=1,\cdots,N\}$ as follows:
\begin{eqnarray}
Q(\lambda)=\prod^{N}_{j=1}(\lambda-\lambda_j).
\end{eqnarray}
To ensure $\Lambda(\lambda)$ is a polynomial, the residues of $\Lambda(\lambda)$ at the
poles $\lambda_j$ must vanish. So the $N$ Bethe roots satisfy the Bethe equations
\begin{eqnarray}
\omega\prod^N_{k=1}\frac{\lambda_j-\theta_k+\eta}{\lambda_j-\theta_k}
=\prod^N_{l\neq j\atop{l=1}}\frac{\lambda_j-\lambda_l+\eta}{\lambda_j-\lambda_l-\eta},\; j=1,\hspace{-0.5mm}\cdots\hspace{-0.5mm},N.~~
\label{BAEs}\end{eqnarray}

Then the energy spectrum of the Hamiltonian (Eq.\ref{Hamiltonian}) can be given in terms of the associated Bethe roots by
\begin{eqnarray}
E&=&-\eta\frac{\partial\ln\Lambda(\lambda)}{\partial \lambda}\bigg{|}_{\lambda=0,\{\theta_j=0\}}+N\nonumber\\
&=&-\sum^N_{j=1}\frac{\eta^2}{\lambda_j(\lambda_j+\eta)}
\end{eqnarray}

\section{Conclusions}
In this paper, we have studied the free Motzkin chain with periodic boundary conditions. Based on  intrinsic properties of the $R$-matrix, we derive the functional relations (Eq.\ref{Lambda-identity}) of the transfer matrix. These relations, together with the asymptotic behavior (Eq.\ref{Lambda-asymptotic}), allow us to construct a generalized $T-Q$ relation (Eq.\ref{T-Q}) for the eigenvalues of the transfer matrix and the associated Bethe equations (Eq.\ref{BAEs}).
The new parameter $\{\omega\}$ in the $T-Q$ relation is related to the roots of unity. Its distribution can be described by the M\"obius function of the number theory. Some patterns in number theory can be observed from the distribution (see equation  (Eq.\ref{Formula_Mobius}), also the Table \ref{tab} in Appendix).
For example, when the lattice site $N$ is a prime number, the total numbers of the square roots and the highest $N$-th roots are both $2^N-2$. Since the degeneracies of all kinds of $N$-th roots are equal, the number $2^N-2$ is divisible by $N$, for any prime $N$.
The exactly solutions allow one to study the thermodynamic properties. While now the Bethe equations have new parameters $\{\omega\}$, they have different values in different sub-spaces. One can not directly apply the Yang-Yang thermodynamic to this model.
Our method may apply to other  Temperley-Lieb  models with crossing unitarity breaking\footnote{We call the free Motzkin $R$-matrix as the $su(3)$ like type. We have generalized the current $R$-matrix without crossing unitarity to higher rank ($su(4)$, $su(5)$ like) models.}.

\section*{Acknowledgments}
KH would like to thank Profs. Yupeng Wang, Wen-Li Yang, Guang-Liang Li, and Junpeng Cao for the helpful discussions.
The work of KH was supported by the National Natural Science Foundation
of China (Grant Nos. 12275214, 11805152, 12047502 and 11947301), the Natural Science Basic Research Program of Shaanxi Province Grant Nos. 2021JCW-19 and 2019JQ-107, and Shaanxi Key Laboratory for Theoretical Physics Frontiers in China.
VK was supported by the SUNY center for QIS at Long Island project number CSP181035.
The authors would like to thank the Simons Center for Geometry and Physics for hospitality.
This work was started during the workshop
``Entanglement and Dynamical Systems: December 10-14, 2018'' held at the center.



\section*{Appendix A: Structure of operator $\hat{O}$ and its eigenvalues}
\setcounter{equation}{0}
\renewcommand{\theequation}{A.\arabic{equation}}

Operator $\hat{O}$ is composed of the leading terms of operators $A(\lambda)$ and $D_{22}(\lambda)$ as $\lambda\rightarrow\infty$.
To study this, besides the local operators $I^u$ in (Eq.\ref{def-u}) and $I^d$ in (Eq.\ref{def-d}),we further introduce the following notations,
\begin{equation}
S^+ =\left|u\right>\left<d\right|=
\left(\begin{array}{ccc}
 0 & 0 & 1 \\
 0 & 0 & 0 \\
 0 & 0 & 0
\end{array}\right),\;
S^- =\left|d\right>\left<u\right|=
\left(\begin{array}{ccc}
 0 & 0 & 0 \\
 0 & 0 & 0 \\
 1 & 0 & 0
\end{array}\right).
\end{equation}
For total lattice sites number $N$, let us set
\begin{eqnarray}
A_{\langle N\rangle}\equiv\lim_{\lambda\rightarrow\infty}\frac{A(\lambda)}{\lambda^N},\;
B_{\langle N\rangle}\equiv\lim_{\lambda\rightarrow\infty}\frac{B_2(\lambda)}{\lambda^N},\;
C_{\langle N\rangle}\equiv\lim_{\lambda\rightarrow\infty}\frac{C_2(\lambda)}{\lambda^N},\;
D_{\langle N\rangle}\equiv\lim_{\lambda\rightarrow\infty}\frac{D_{22}(\lambda)}{\lambda^N}.
\end{eqnarray}
Here for $B_2(\lambda)$, the subscript $2$ refers to the second $B$ type operator in the main text (Eq.\ref{T-matrix}).
While the subscript of $B_{\langle N\rangle}$ is the number of total system sites, and these operators act nontrivially on all the $N$ quantum spaces. Then for the system with $N$ lattice sites,
\begin{eqnarray}
\hat{O}_{\langle N\rangle}=A_{\langle N\rangle}+D_{\langle N\rangle}.
\end{eqnarray}
Because of the symmetry between $A(\lambda)$ and $D_{22}(\lambda)$, we only need to analyze the structure of operator $A(\lambda)$.
According to the structure of the monodromy matrix and matrix multiplication, we have the following recursive relations for these operators:
\begin{eqnarray}
&A_{\langle1\rangle}=I^u_1,\quad B_{\langle1\rangle}=S^-_1, \nonumber\\
&A_{\langle N\rangle}=A_{\langle N-1\rangle}\otimes I^u_N +B_{\langle N-1\rangle}\otimes S^+_N,\quad
B_{\langle N\rangle}=A_{\langle N-1\rangle}\otimes S^-_N+B_{\langle N-1\rangle}\otimes I^d_N.
\end{eqnarray}
Subscripts of $I^u$, $I^d$, $S^+$ and $S^-$ refer to the quantum spaces on which these local operators act.
Based on the above recursive relations, we have the following rules for the structure of the operator $A_{\langle N\rangle}$:
\begin{itemize}
\item The operator $A_{\langle N\rangle}$ is composed of local operators $I^u$, $I^d$, $S^-$ and $S^+$.
\item $A_{\langle N\rangle}$ is the sum of products of $N$ local operators acting on each of the quantum spaces(from $1$ to $N$, respectively).
\item $A_{\langle N\rangle}$ is the sum of all possible operator products permutations(terms).
\item $S^-$ and $S^+$ operators appear in pairs, and $S^-$ always comes before $S^+$ in one permutation(term). That means the quantum subspace number of $S^-$ is smaller than that of $S^+$.
\item $I^u$ can appear on its own. It can also come before $S^-$ or after $S^+$. $I^d$ only appears between $S^-$ and $S^+$.
\item Multiple $S^-S^+$ pairs appear in turns. They can not be nested within each other.
\end{itemize}

\subsection*{Appendix A1: For $\hat{O}_{\langle3\rangle}$, $\hat{O}_{\langle6\rangle}$ and $\hat{O}_{\langle7\rangle}$ cases, the structure and the rules of the eigenvalue distribution}

In the following, we take $N=3$ as an example. $A_{\langle3\rangle}$ can be represented as
\begin{eqnarray}
A_{\langle3\rangle}=I^u_1I^u_2I^u_3+I^u_1S^-_2S^+_3+S^-_1S^+_2I^u_3+S^-_1I^d_2S^+_3.
\end{eqnarray}
Considering the symmetry between $A_{\langle3\rangle}$ and $D_{\langle3\rangle}$, the operator $\hat{O}_{\langle3\rangle}=A_{\langle3\rangle}+D_{\langle3\rangle}$ has the matrix form

\begin{equation}
\left(
  \begin{array}{ccccccccc|ccccccccc|ccccccccc}
1&0&0&0&0&0&0&0&0&0&0&0&0&0&0&0&0&0&0&0&0&0&0&0&0&0&0\\
0&1&0&0&0&0&0&0&0&0&0&0&0&0&0&0&0&0&0&0&0&0&0&0&0&0&0\\
0&0&0&0&0&0&0&0&0&0&0&0&0&0&0&0&0&0&\color{blue}{1}&0&0&0&0&0&0&0&0\\
0&0&0&1&0&0&0&0&0&0&0&0&0&0&0&0&0&0&0&0&0&0&0&0&0&0&0\\
0&0&0&0&1&0&0&0&0&0&0&0&0&0&0&0&0&0&0&0&0&0&0&0&0&0&0\\
0&0&0&0&0&0&0&0&0&0&0&0&0&0&0&0&0&0&0&0&0&\color{red}{1}&0&0&0&0&0\\
0&0&\color{blue}{1}&0&0&0&0&0&0&0&0&0&0&0&0&0&0&0&0&0&0&0&0&0&0&0&0\\
0&0&0&0&0&0&0&0&0&0&0&0&0&0&0&0&0&0&0&\color{red}{1}&0&0&0&0&0&0&0\\
0&0&0&0&0&0&0&0&0&0&0&0&0&0&0&0&0&0&0&0&\color{blue}{1}&0&0&0&0&0&0\\
\hline
0&0&0&0&0&0&0&0&0&1&0&0&0&0&0&0&0&0&0&0&0&0&0&0&0&0&0\\
0&0&0&0&0&0&0&0&0&0&1&0&0&0&0&0&0&0&0&0&0&0&0&0&0&0&0\\
0&0&0&0&0&0&0&0&0&0&0&0&0&0&0&\color{red}{1}&0&0&0&0&0&0&0&0&0&0&0\\
0&0&0&0&0&0&0&0&0&0&0&0&\textbf{1}&\textbf{0}&\textbf{0}&0&0&0&0&0&0&0&0&0&0&0&0\\
0&0&0&0&0&0&0&0&0&0&0&0&\textbf{0}&\textbf{2}&\textbf{0}&0&0&0&0&0&0&0&0&0&0&0&0\\
0&0&0&0&0&0&0&0&0&0&0&0&\textbf{0}&\textbf{0}&\textbf{1}&0&0&0&0&0&0&0&0&0&0&0&0\\
0&0&0&0&0&0&0&0&0&0&0&\color{red}{1}&0&0&0&0&0&0&0&0&0&0&0&0&0&0&0\\
0&0&0&0&0&0&0&0&0&0&0&0&0&0&0&0&1&0&0&0&0&0&0&0&0&0&0\\
0&0&0&0&0&0&0&0&0&0&0&0&0&0&0&0&0&1&0&0&0&0&0&0&0&0&0\\
\hline
0&0&0&0&0&0&\color{blue}{1}&0&0&0&0&0&0&0&0&0&0&0&0&0&0&0&0&0&0&0&0\\
0&0&0&0&0&0&0&\color{red}{1}&0&0&0&0&0&0&0&0&0&0&0&0&0&0&0&0&0&0&0\\
0&0&0&0&0&0&0&0&0&0&0&0&0&0&0&0&0&0&0&0&0&0&0&0&\color{blue}{1}&0&0\\
0&0&0&0&0&\color{red}{1}&0&0&0&0&0&0&0&0&0&0&0&0&0&0&0&0&0&0&0&0&0\\
0&0&0&0&0&0&0&0&0&0&0&0&0&0&0&0&0&0&0&0&0&0&1&0&0&0&0\\
0&0&0&0&0&0&0&0&0&0&0&0&0&0&0&0&0&0&0&0&0&0&0&1&0&0&0\\
0&0&0&0&0&0&0&0&\color{blue}{1}&0&0&0&0&0&0&0&0&0&0&0&0&0&0&0&0&0&0\\
0&0&0&0&0&0&0&0&0&0&0&0&0&0&0&0&0&0&0&0&0&0&0&0&0&1&0\\
0&0&0&0&0&0&0&0&0&0&0&0&0&0&0&0&0&0&0&0&0&0&0&0&0&0&1\\
  \end{array}
\right).
\end{equation}

In this matrix, $A_{\langle3\rangle}$ occupies the left $14$ columns, and $D_{\langle3\rangle}$ occupies the right $14$ columns. The number $2=1+1$ in the center (the entry in the $14$-th row and $14$-th column, $2\times E^{MM}$, $M=\frac{3^3+1}{2}=14$) is the sum of value $1$ from $A_{\langle3\rangle}$ and another $1$ from $D_{\langle3\rangle}$.

In this matrix form, we divide the first quantum space into $3\times3$ blocks by solid lines. The block in the center, $|{f_1}\rangle\langle{f_1}|=E^{2,2}_1$ (we adopt the standard basis, subscript refers to quantum space) is the operator $\hat{O}_{\langle2\rangle}$. The boldface $3\times3$ diagonal matrix $\operatorname {diag}\{1,2,1\}$ in the middle is just the operator $\hat{O}_{\langle1\rangle}$. Then the $2$ in the center can also be represented as $2\times E^{MM}=2\times E^{2,2}_1E^{2,2}_2E^{2,2}_3$ in quantum subspaces.

We can analyze the distribution of eigenvalues of $\hat{O}_{\langle3\rangle}$ from operator $A_{\langle3\rangle}$. The term $I^u_1I^u_2I^u_3$ corresponds to all the diagonal elements. Considering the symmetry between $A_{\langle3\rangle}$ and $D_{\langle3\rangle}$, the operator $\hat{O}_{\langle3\rangle}$ has $2^4-1=15$ diagonal elements, including one single $2$, and $14$ elements equal to $1$.

Square roots of unity are caused by $S^-S^+$ pairs. Each term of $A_{\langle3\rangle}$ expression consists of $3$ local operators. In each term with a $S^-S^+$ pair, the center element $E^{2,2}$ (in its own subspace) of the rest operator $I^u$(or $I^d$) will lead to a square root.
In $A_{\langle3\rangle}$ operator, these elements are $E^{2,2}_1S^-_2S^+_3$, $S^-_1S^+_2E^{2,2}_3$ and $S^-_1E^{2,2}_2S^+_3$. Considering $D_{\langle3\rangle}$, there are $6$ elements related to square roots in $\hat{O}_{\langle3\rangle}$. This number can be calculated by combination\footnote{For brevity, we adopt the inline notation of the binomial coefficient $C(n,k)=\binom {n}{k}$.} $C(3,2)\times2=6$. The $6$ elements are marked with red color in the above matrix. These elements will lead to $6$ square roots of unity: $-1$, $-1$, $-1$, $1$, $1$ and $1$.

The $S^-S^+$ pairs are also the basic components of higher order roots of unity.

Cubic roots come from terms with one $S^-S^+$ pair. Unlike the case of square roots, in terms with a $S^-S^+$ pair ($3$ local operators in each term), the rest one operator $I^u$(or $I^d$) should take $E^{1,1}$(or $E^{3,3}$). This $E^{1,1}$ (or $E^{3,3}$) makes the order plus one: $2+1=3$. Then in $A_{\langle3\rangle}$ operator, these elements are $E^{1,1}_1S^-_2S^+_3$, $S^-_1S^+_2E^{1,1}_3$ and $S^-_1E^{3,3}_2S^+_3$. Considering $D_{\langle3\rangle}$, there are $6$ elements in $\hat{O}_{\langle3\rangle}$. We calculate the number as $C(3,2)\times C(1,1)\times2=6$.
As marked in blue in the $\hat{O}_{\langle3\rangle}$ matrix form, the $6$ elements will lead to $6$ cubic roots: $\exp({2\pi i\over3})$, $\exp({2\pi i\over3})$, $\exp({4\pi i\over3})$, $\exp({4\pi i\over3})$, $1$ and $1$.

Next let us take operator $A_{\langle6\rangle}$ as an example to explain more rules of eigenvalues of $\hat{O}$. The operator $A_{\langle6\rangle}$ can be expressed as the sum of the products of $6$ local operators
\begin{eqnarray}
A_{\langle6\rangle}&=&I^u_1 I^u_2 I^u_3 I^u_4 I^u_5 I^u_6
+I^u_1 I^u_2 I^u_3 I^u_4 S^{-}_5 S^{+}_6
+I^u_1 I^u_2 I^u_3 S^{-}_4 S^{+}_5 I^u_6
+I^u_1 I^u_2 S^{-}_3 S^{+}_4 I^u_5 I^u_6
\no\\
&+&I^u_1 S^{-}_2 S^{+}_3 I^u_4 I^u_5 I^u_6
+S^{-}_1 S^{+}_2 I^u_3 I^u_4 I^u_5 I^u_6
+I^u_1 I^u_2 I^u_3 S^{-}_4 I^d_5 S^{+}_6
+I^u_1 I^u_2 S^{-}_3 I^d_4 S^{+}_5 I^u_6
\no\\
&+&I^u_1 S^{-}_2 I^d_3 S^{+}_4 I^u_5 I^u_6
+S^{-}_1 I^d_2 S^{+}_3 I^u_4 I^u_5 I^u_6
+I^u_1 I^u_2 S^{-}_3 I^d_4 I^d_5 S^{+}_6
+I^u_1 S^{-}_2 I^d_3 I^d_4 S^{+}_5 I^u_6
\no\\
&+&S^{-}_1 I^d_2 I^d_3 S^{+}_4 I^u_5 I^u_6
+I^u_1 S^{-}_2 I^d_3 I^d_4 I^d_5 S^{+}_6
+S^{-}_1 I^d_2 I^d_3 I^d_4 S^{+}_5 I^u_6
+S^{-}_1 I^d_2 I^d_3 I^d_4 I^d_5 S^{+}_6
\no\\
&+&I^u_1 I^u_2 S^{-}_3 S^{+}_4 S^{-}_5 S^{+}_6
+S^{-}_1 S^{+}_2 I^u_3 I^u_4 S^{-}_5 S^{+}_6
+S^{-}_1 S^{+}_2 S^{-}_3 S^{+}_4 I^u_5 I^u_6
\no\\
&+&I^u_1 S^{-}_2 S^{+}_3 I^u_4 S^{-}_5 S^{+}_6
+I^u_1 S^{-}_2 S^{+}_3 S^{-}_4 S^{+}_5 I^u_6
+S^{-}_1 S^{+}_2 I^u_3 S^{-}_4 S^{+}_5 I^u_6
\no\\
&+&I^u_1 S^{-}_2 I^d_3 S^{+}_4 S^{-}_5 S^{+}_6
+I^u_1 S^{-}_2 S^{+}_3 S^{-}_4 I^d_5 S^{+}_6
+S^{-}_1 S^{+}_2 I^u_3 S^{-}_4 I^d_5 S^{+}_6
+S^{-}_1 I^d_2 S^{+}_3 I^u_4 S^{-}_5 S^{+}_6
\no\\
&+&S^{-}_1 S^{+}_2 S^{-}_3 I^d_4 S^{+}_5 I^u_6
+S^{-}_1 I^d_2 S^{+}_3 S^{-}_4 S^{+}_5 I^u_6
\no\\
&+&S^{-}_1 S^{+}_2 S^{-}_3 I^d_4 I^d_5 S^{+}_6
+S^{-}_1 I^d_2 S^{+}_3 S^{-}_4 I^d_5 S^{+}_6
+S^{-}_1 I^d_2 I^d_3 S^{+}_4 S^{-}_5 S^{+}_6
\no\\
&+&S^{-}_1 S^{+}_2 S^{-}_3 S^{+}_4 S^{-}_5 S^{+}_6
\end{eqnarray}
Among them, the term $I^u_1I^u_2I^u_3I^u_4I^u_5I^u_6$ corresponds to all the diagonal elements. Similarly, together with $D_{\langle6\rangle}$, there are $2^7-1=127$ diagonal elements in $\hat{O}_{\langle6\rangle}$, and one of them is $2$. All the rest of the elements are $1$.

The number of square roots of unity is $(C(6,2)+C(6,4)+C(6,6))\times2=2^6-2=62$, and half ($62/2=31$) of them are $-1$, the other half are $1$. Single $S^-S^+$ pair in one term leads to square roots, when all the rest operators take $E^{2,2}$. Recurring $S^-S^+$ pairs ($2$ or $3$) in one term with the rest operators taking $E^{2,2}$, will also give rise to square roots. This means we should also take terms with $S^-S^+S^-S^+$ and $S^-S^+S^-S^+S^-S^+$ into account.

The number of cubic roots is $(C(6,2)\times C(4,1)+C(3,1))\times2=126$. The three kinds of cubic roots are $\exp({2\pi i\over3})$, $\exp({4\pi i\over3})$ and $1$, and each has degeneracy $126\over3$.
In the expression of operator $A_{\langle6\rangle}$, let us consider terms with one $S^-S^+$ pair and four $I^u$(or $I^d$) operators. When one $I^u$(or $I^d$) operator takes $E^{1,1}$(or $E^{3,3}$) in its subspace, and the other three $I^u$ take $E^{2,2}$, each such choice in one term will lead to a cubic root. And recurring $I^uS^-S^+$($I^u$ takes $E^{11}$), $S^-I^dS^+$($I^d$ takes $E^{33}$) and $S^-S^+I^u$($I^u$ takes $E^{11}$) patterns in one term also lead to cubic roots.

The number of $4$-th roots is $(C(6,2)\times C(4,2))\times2=180$. The four kinds of $4$-th roots are $i$, $-1$, $-i$ and $1$, and each kind has degeneracy $180\over4$.
$4$-th roots come from terms with a $S^-S^+$ pair and two $I^u$(or $I^d$) taking $E^{1,1}$(or $E^{3,3}$) in its subspace, and the other two $I^u$(or $I^d$) taking $E^{2,2}$. Two $S^-S^+$ pairs in one term with the rest two operators taking $E^{2,2}$, will reduce to square roots, as we mentioned above.

The number of $5$-th roots is $(C(6,2)\times C(4,3)+C(6,4)\times C(2,1))\times2=180$. The five kinds of $5$-th roots are $\exp({2\pi i\over5})$, $\exp({4\pi i\over5})$, $\exp({6\pi i\over5})$, $\exp({8\pi i\over5})$ and $1$, and each kind has degeneracy $180\over5$.
$5$-th roots come from terms with a $S^-S^+$ pair and three $I^u$(or $I^d$) taking $E^{1,1}$(or $E^{3,3}$) in its subspace, and the other one $I^u$(or $I^d$) taking $E^{2,2}$.
$5$-th roots also come from terms with two $S^-S^+$ pairs, and one of the rest two operators $I^u$(or $I^d$) taking $E^{1,1}$(or $E^{3,3}$) in its subspace, and the other taking $E^{2,2}$.

The number of $6$-th roots is $(C(6,2)\times C(4,4)+C(6,4)\times C(2,2)- C(3,1))\times2=54$.
The six kinds of $6$-th roots are $\exp({2\pi i\over6})$, $\exp({4\pi i\over6})$, $-1$, $\exp({8\pi i\over6})$, $\exp({10\pi i\over6})$ and $1$, each kind has degeneracy $54\over6$.
$6$-th roots come from terms with all operators not taking their center element $E^{2,2}$ in their subspace. This will contain terms such as $S^{-}_1S^{+}_2E^{1,1}_3E^{1,1}_4E^{1,1}_5E^{1,1}_6$ and $S^{-}_1S^{+}_2S^{-}_3S^{+}_4E^{1,1}_5E^{1,1}_6$.
While there are exceptions, recurring $I^uS^-S^+$, $S^-I^dS^+$ and $S^-S^+I^u$ pairs such as $E^{1,1}_1S^-_2S^+_3E^{1,1}_4S^-_5S^+_6$, $S^-_1E^{3,3}_2S^+_3S^-_4E^{3,3}_5S^+_6$ and $S^-_1S^+_2E^{1,1}_3S^-_4S^+_5E^{1,1}_6$ will reduce to cubic roots. $S_1^-S_2^+S_3^-S_4^+S_5^-S_6^+$ will reduce to square root (as mentioned above).

For the case of $\hat{O}_{\langle7\rangle}$, according to the above rules, we directly write down the distribution of different eigenvalues.

Eigenvalues from diagonal elements: a single $2$, multiple $1$ with degeneracy of $2^8-2=254$;

Number of square roots of unity: $(C(7,2)+C(7,4)+C(7,6))\times2=2^7-2=126$. $-1$ and $1$ are roots of degeneracy ${126\over2}=63$.

Number of cubic roots: $(C(7,2)\times C(5,1)+ C(3,1)\times7)\times2=252$. $\exp({2\pi i\over3})$, $\exp({4\pi i\over3})$ and $1$ are roots of degeneracy ${252\over3}=84$.

Number of $4$-th roots: $(C(7,2)\times C(5,2))\times2=420$. $i$, $-1$, $-i$ and $1$ are roots of degeneracy ${420\over4}=105$.

Number of $5$-th roots: $(C(7,2)\times C(5,3)+C(7,4)\times C(3,1))\times2=630$. $\exp({2\pi i\over5})$, $\exp({4\pi i\over5})$, $\exp({6\pi i\over5})$, $\exp({8\pi i\over5})$ and $1$ are roots of degeneracy ${630\over5}=126$.

Number of $6$-th roots: $(C(7,2)\times C(5,4)+C(7,4)\times C(3,2)-C(3,1)\times7)\times2=378$. $\exp({2\pi i\over6})$, $\exp({4\pi i\over6})$, $-1$, $\exp({8\pi i\over6})$, $\exp({10\pi i\over6})$ and $1$ are roots of degeneracy ${378\over6}=63$.

Number of $7$-th roots: $(C(7,2)\times C(5,5)+C(7,4)\times C(3,3)+C(7,6)\times C(1,1))\times2=126$. $\exp({2\pi i\over7})$, $\exp({4\pi i\over7})$, $\exp({6\pi i\over7})$, $\exp({8\pi i\over7})$, $\exp({10\pi i\over7})$, $\exp({12\pi i\over7})$ and $1$ are roots of degeneracy ${126\over7}=18$.

\subsection*{Appendix A2: The combination rules for the eigenvalues of operator $\hat{O}_{\langle N\rangle}$}

Further, we summarize the combination rules for the eigenvalues of operator $\hat{O}_{\langle N\rangle}$:
\begin{itemize}
\item Number of diagonal elements is $2^{N+1}-1$, and one of them is $2$. All the rest are $1$.
\item Number of square roots is $C(N,2)+C(N,4)+\cdots=2^N-2$. Square roots come from $S^-S^+$ pair(s), and recurring $S^-S^+$ also give rise to square roots.
\item Higher order roots are based on $S^-S^+$ pair(s). In one term of operator products, each element $E^{1,1}$ in $I^u$(or $E^{3,3}$ in $I^d$) will increase the order by $1$, and each additional $S^-S^+$ pair will increase the order by $2$. The element $|{f}\rangle\langle{f}|=E^{2,2}$ in the center of each subspace will maintain the order number. Briefly speaking, in a certain term, each element in local operator outside the center of its subspace\footnote{When we talk about this rule, we exclude the term with all local operators being $I^u$, namely $I^u_1\cdots I^u_N$ which correspond to all diagonal elements.} makes the order of the root plus $1$. In contrast, elements in the center $E^{2,2}$ of local operators ($I^u$ and $I^d$) have no contribution to the order.
\item Terms with recurring higher order patterns (like $S^-S^+$ is related to square root and $S^-S^+E^{1,1}$ is related to cubic root) will reduce their order to a single pattern. When counting $n$-th order terms, if $n$ is a composite number, we should consider all kinds of recurring patterns (which means all divisors of $n$) that lead to reduced orders.
\item In every specific $m$-th roots case, each kind of roots $\exp({2\pi i\over m})$, $\exp({4\pi i\over m})$,$\cdots$, $1$ has the same degeneracy.
\end{itemize}

\section*{Appendix B: The distribution formula for the eigenvalues of operator $\hat{O}$}
\setcounter{equation}{0}
\renewcommand{\theequation}{B.\arabic{equation}}

Let us recall the definition of $\hat{O}$. It is the leading term in the taylor expansion of the transfer matrix, i.e. $t(\lambda)= A(\lambda) + D_{11}(\lambda) + D_{22}(\lambda) = (\mathbbold{1} + \hat{O})\lambda^N +\cdots$ Equivalently, it may be defined as:
\begin{eqnarray}
\mathbbold{1} + \hat{O} = \text{tr} \{ \Tilde{R}_{0N} \Tilde{R}_{0N-1} ... \Tilde{R}_{02} \Tilde{R}_{01} \}
\end{eqnarray}
where $\Tilde{R} = \lim_{\lambda \rightarrow \infty} \frac{R(\lambda)}{\lambda} = \mathbbold{1}  - (|{ud}\rangle - |{du}\rangle)(\langle{ud}| - \langle{du}|)$ can be seen as a permutation matrix which permutes the states $|{u}\rangle \otimes |{d}\rangle$ and $|{d}\rangle \otimes |{u}\rangle$. One immediate consequence of this is that the leading term of $D_{11}$ is $\mathbbold{1}$ (since the auxiliary space is left invariant), which we chose to cancel out with the $\mathbbold{1}$ term in our definition.

The second consequence is that the operator maps basis states to basis states. Apart from the state $|{f}\rangle \otimes|{f}\rangle \otimes ... \otimes |{f}\rangle \otimes |{f}\rangle$ which is mapped to itself by both $A_{\langle{N}\rangle}$ and $D_{\langle{N}\rangle}$ and so is an eigenstate with eigenvalue 2, the operator acts as a permutation matrix on the basis states. It acts as a left cyclic shift of all sites not equal to $|{f}\rangle$. For example, on a chain of length $3$ it maps $\ket{u} \otimes\ket{d}\otimes\ket{u}$ to $\ket{d} \otimes\ket{u}\otimes\ket{u}$.
On a chain of length $6$, the state $|{u}\rangle \otimes |{u}\rangle \otimes|{d}\rangle \otimes|{f}\rangle \otimes|{u}\rangle \otimes|{d}\rangle$ can be mapped to  $|{u}\rangle \otimes |{d}\rangle \otimes|{u}\rangle \otimes|{f}\rangle \otimes|{d}\rangle \otimes|{u}\rangle$.

As a result, finding the spectrum of $\Hat{O}$ is equivalent to finding the cycle decomposition of the corresponding permutation. This is also equivalent to the combination rules in the last section. We can write down the distribution of eigenvalues for arbitrary $\hat{O}_{\langle N\rangle}$.

The number of $k$-th roots for a lattice of length $N$ is given by:
\begin{eqnarray}
f(N,k) = \left( \sum_{d|k} 2^d \mu(\frac{k}{d}) \right) \left[ \sum_{m = 1}^{\lfloor{\frac{N}{k}}\rfloor} C(N,km)
\right]
\end{eqnarray}
where $\mu(n)$ refers to the M\"obius function. For $k = 1$ this formula gives the correct diagonal result if one excludes the eigenvalue $2$.

Some results in number theory can be observed from the distribution. For example, when $N$ is a prime number, the total number of square roots is $2^N-2$. It is equal to the total number of the highest $N$-th roots in this case. As we mentioned above, the degeneracies of all kinds of $N$-th roots are equal. This will lead to a result in number theory: $2^N-2$ is divisible by $N$, for $N$ is prime.

We can tabulate the first few values, which gives in Table \ref{tab}:
\begin{table}[H]
    \centering
    \begin{tabular}{ccccccccccc}
Length & Diag & sqrt & 3rd & 4th & 5th & 6th & 7th & 8th & 9th & 10th \\
\hline
 N = 1 & 2 & 0 & 0 & 0 & 0 & 0 & 0 & 0 & 0 & 0 \\
 N = 2 & 6 & 2 & 0 & 0 & 0 & 0 & 0 & 0 & 0 & 0 \\
 N = 3 & 14 & 6 & 6 & 0 & 0 & 0 & 0 & 0 & 0 & 0 \\
 N = 4 & 30 & 14 & 24 & 12 & 0 & 0 & 0 & 0 & 0 & 0 \\
 N = 5 & 62 & 30 & 60 & 60 & 30 & 0 & 0 & 0 & 0 & 0 \\
 N = 6 & 126 & 62 & 126 & 180 & 180 & 54 & 0 & 0 & 0 & 0 \\
 N = 7 & 254 & 126 & 252 & 420 & 630 & 378 & 126 & 0 & 0 & 0 \\
 N = 8 & 510 & 254 & 504 & 852 & 1680 & 1512 & 1008 & 240 & 0 & 0 \\
 N = 9  & 1022 & 510 & 1014 & 1620 & 3780 & 4536 & 4536 & 2160 & 504 & 0 \\
 N = 10 &  2046 & 1022 & 2040 & 3060 & 7590 & 11340 & 15120 & 10800 & 5040 & 990 \\
\hline
\end{tabular}
\caption{The distribution of $k$-th roots of unity for a lattice of length $N$. The categories and quantities of the eigenvalues for the operator $\hat{O}$ from $N=1$ to $10$. The results from direct diagonalization are coincident with the numbers we count from the combination rules $f(N,k)$. We remark that in the diagonal elements, the eigenvalue $2$ is excluded, and $1$ can be regarded as the $1$-st root of unity.}
    \label{tab}
\end{table}
This was computed with the following implementation in Mathematica:
\begin{verbatim}
  g[n_]:= DivisorSum[n,(2^#)*MoebiusMu[n/#]&]
  binomSum[n_,k_]:=Sum[Binomial[n, i],{i,k,n,k}]

  f[n_,k_]:=g[k]*binomSum[n,k]
\end{verbatim}

\providecommand{\noopsort}[1]{}\providecommand{\singleletter}[1]{#1}%

\end{document}